\documentclass[conference]{IEEEtran}
\IEEEoverridecommandlockouts
\usepackage{cite}
\usepackage{amsmath,amssymb,amsfonts}
\usepackage{algorithmic}
\usepackage{graphicx}
\usepackage{textcomp}
\usepackage{xcolor}
\usepackage{enumitem}
\usepackage{array}
\usepackage{soul}
\usepackage{subcaption}
\usepackage[font=small]{caption}
\newcommand{\PreserveBackslash}[1]{\let\temp=\\#1\let\\=\temp}
\newcolumntype{C}[1]{>{\PreserveBackslash\centering}p{#1}}
\newcolumntype{R}[1]{>{\PreserveBackslash\raggedleft}p{#1}}
\newcolumntype{L}[1]{>{\PreserveBackslash\raggedright}p{#1}}

\def\BibTeX{{\rm B\kern-.05em{\sc i\kern-.025em b}\kern-.08em
    T\kern-.1667em\lower.7ex\hbox{E}\kern-.125emX}}
\begin{document}

%\title{Neighbors From Hell: On Multi-Tenant FPGA Voltage Attacks and the Resilience of DL Models}
\title{Neighbors From Hell: Voltage Attacks Against Deep Learning Accelerators on Multi-Tenant FPGAs}

\author{
\IEEEauthorblockN{Andrew Boutros$^{1,2}$, Mathew Hall$^{1}$, Nicolas Papernot$^{1,2}$ and Vaughn Betz$^{1,2}$}
\IEEEauthorblockA{$^{1}$Department of Electrical and Computer Engineering, University of Toronto, Toronto, ON, Canada \\
$^{2}$Vector Institute, Toronto, ON, Canada\\
\{andrew.boutros, mathew.hall\}@mail.utoronto.ca, nicolas.papernot@utoronto.ca, vaughn@eecg.utoronto.ca}
\vspace{-1.1cm}
}

\maketitle

\begin{abstract}
Field-programmable gate arrays (FPGAs) are becoming widely used accelerators for a myriad of datacenter applications due to their flexibility and energy efficiency.
Among these applications, FPGAs have shown promising results in accelerating low-latency real-time deep learning (DL) inference, which is becoming an indispensable component of many end-user applications.
With the emerging research direction towards \emph{virtualized} cloud FPGAs that can be shared by multiple users, the security aspect of FPGA-based DL accelerators requires careful consideration.
In this work, we evaluate the security of DL accelerators against voltage-based integrity attacks in a multi-tenant FPGA scenario.
We first demonstrate the feasibility of such attacks on a state-of-the-art Stratix 10 card using different attacker circuits that are logically and physically isolated in a separate \emph{attacker} role, and cannot be flagged as malicious circuits by conventional bitstream checkers.
We show that aggressive clock gating, an effective power-saving technique, can also be a potential security threat in modern FPGAs. 
Then, we carry out the attack on a DL accelerator running ImageNet classification in the \emph{victim} role to evaluate the inherent resilience of DL models against timing faults induced by the adversary.
We find that, even when using the strongest attacker circuit, the prediction accuracy of the DL accelerator is not compromised when running at its safe operating frequency. 
Furthermore, we can achieve 1.18$-$1.31$\times$ higher inference performance by over-clocking the DL accelerator without affecting its prediction accuracy.
\end{abstract}


\fontsize{9.5pt}{11.5pt}\selectfont

\vspace{-0.6cm}
\section{Introduction}
% FPGAs are being widely deployed in datacenters and used for DL inference
Field-programmable gate arrays (FPGAs) are being deployed on a large scale in datacenters due to their flexibility and energy efficiency.
The Catapult project \cite{putnam2014reconfigurable} coupled every server node in Microsoft's datacenters with an FPGA to accelerate search engines, and perform on-the-fly compression and cryptography for the data transferred over the network \cite{caulfield2016cloud}.
Currently, FPGAs in the cloud are either used by a service provider to offload specific datacenter workloads from CPUs, as in the case of Microsoft's Catapult project \cite{putnam2014reconfigurable}, or rented out to external users as single-tenant compute nodes, as in Amazon F1 instances that come with 1, 2 or 8 FPGAs \cite{amazon}.
% Multi-Tenant FPGAs
However, several studies envision FPGAs in the cloud as \emph{virtualized} compute resources that can be shared by multiple users, similar to the traditional datacenter CPUs \cite{yazdanshenas2018interconnect}.
This requires abstracting the low-level design details and external interfaces of an FPGA from the user.
Typically, the service provider designs a \emph{shell} that handles all the interfacing to external resources, such as off-chip memory, Ethernet, and PCIe, and delivers the bandwidth of these external interfaces to user \emph{roles}.
In this case, user roles are partially reconfigurable regions that implement user-specified functionality and can be re-programmed without disrupting the operation of other roles.

% Deep Learning is becoming a major datacenter workload
As deep learning (DL) is rapidly becoming the cornerstone of many real-time datacenter services, service providers implement and deploy highly-efficient specialized accelerators to handle the ever-increasing computational demands of DL workloads.
In these DL-based services, such as speech-inquired smart assistants and machine translation, low latency is key for a seamless user experience.
In 2014, the maximum tolerated latency for Google's datacenter DL workloads was 10 ms, which went down to 7 ms in 2016 \cite{jouppi2017datacenter}.
Later in 2017, Microsoft used its FPGA cloud to build a cloud-scale DL inference engine, Brainwave \cite{fowers2018configurable}, targeting the lowest possible inference latency.
A single-node Brainwave implemented on an Intel Stratix 10 FPGA runs DeepBench models in less than 4 ms inference latency \cite{fowers2018configurable}, achieving up to 8.6$\times$ lower latency compared to same generation GPUs \cite{nurvitadhi2019compete}.

With the increasing research focus on the reliability, privacy, and security aspects of DL models \cite{papernot2017practical, papernot2018scalable, elsayed2018adversarial}, we believe there is a big gap in understanding the security implications of DL accelerators, especially on reconfigurable and potentially shared devices as FPGAs.
In this work, we focus on multi-tenant FPGAs in the cloud and we demonstrate the feasibility of voltage attacks on a state-of-the-art Intel Stratix 10 FPGA card in a multi-tenancy scenario (i.e. physically isolated user roles).
In a voltage attack, a malicious \emph{attacker} circuit implements logic that draws a large amount of current, causing voltage drops in the chip's power distribution network (PDN).
These voltage drops can induce timing violations in a neighboring \emph{victim} circuit and potentially lead to faulty functionality.
We first use a simple timing violation detection circuit to quantify the safety timing margins introduced by the FPGA CAD tools.
Then, we characterize the effects of the voltage attack using different attacker circuits, some of which use vendor-supplied clock gating circuitry.
Unlike prior work, these attacker circuits can not be detected by conventional bitstream verifiers as malicious circuits. 
In fact, designs using clock gating as a power-saving mechanism could unintentionally create voltage disruptions and timing violations similar to our attacker circuits.
We finally study the effect of the voltage attack on the integrity of a victim role implementing a state-of-the-art DL accelerator. 
%We assume a \emph{black-box} attack in which the adversary does not know any details about either the model or the accelerator architecture in the victim role.

\noindent
To the best of our knowledge, this work is the first attempt to:
\begin{itemize}[noitemsep,topsep=0pt,leftmargin=2\labelsep]
    \item demonstrate voltage attacks on a modern Intel Stratix 10 FPGA in a multi-tenant scenario with unaltered placement and routing of attacker/victim circuits,
    \item investigate the use of clock gating as a potential security threat that cannot be detected by conventional bitstream checkers,
    \item quantify the resilience of DL models against timing violations, introducing the accelerator's operating frequency as an additional knob, similar to model sparsity and precision, that offers a tradeoff between computational efficiency and model accuracy.
\end{itemize}
\section{Background and Related Work}
\label{sec:background}

\subsection{Security of Virtualized FPGAs}
%FPGA Virtualization
With the large-scale deployment of FPGAs in datacenters, multiple previous studies have investigated the virtualization of FPGAs in the cloud.
The terms \emph{shell} and \emph{role} were first defined in \cite{putnam2014reconfigurable} as the portion of reusable FPGA logic common across all applications, and the application logic itself, respectively.
There are multiple industrial \cite{weerasinghe2016network, caulfield2016cloud} and academic \cite{tarafdar2017enabling, byma2014fpgas, yazdanshenas2018interconnect} proposals for the shell implementation, which either connects the FPGA to the host processor through PCIe or place the FPGA as a bump-on-the-wire between the host CPU and the network.
Some of the proposed shell implementations also enable multi-tenancy by supporting up to four application roles per FPGA device \cite{byma2014fpgas, yazdanshenas2018interconnect}. 
In this work, we focus on multi-tenant virtualized FPGAs.
Although we limit our experiments to two roles per device for simplicity, our work is not fundamentally limited to a specific number of roles or a specific shell design.
We point the reader to \cite{vaishnav2018survey} for a comprehensive survey on FPGA virtualization.

Virtualizing FPGAs in the cloud comes with a myriad of security threats that have been investigated in previous studies and can be classified as follows.
This section is intended to place our work in context; however, a detailed survey on security of cloud FPGAs can be found in \cite{jin2020security}.

\subsubsection*{\textbf{Extraction Attacks}}
In this class of attacks, the adversary's main goal is to obtain unauthorized access to data that belongs to the victim circuit.
For example, the authors of \cite{ramesh2018fpga} implement an adversary circuit that can sense the crosstalk from neighbouring long wires of a victim circuit and extract the data transmitted on it. 
They successfully extract the encryption key from an AES encryption core by exploiting this side-channel information leakage.
The long wire data leakage vulnerability is further investigated in \cite{provelengios2019characterization} and found to be exploitable in three different generations of FPGA chips.
This attack requires placing the adversary sensing circuitry adjacent to a victim wire.
Therefore, it assumes a \emph{white-box} attack in which the adversary is aware of the physical design implementation details of the victim circuit and has the ability to impose such placement constraints during the CAD flow.
In \cite{zhao2018fpga}, the authors implement an on-chip power monitoring circuit in the adversary role to observe the power consumption of the victim circuit and use it to perform a power analysis attack and extract the key of an RSA cryptography engine.
Another attack that is mentioned, but not practically demonstrated, in \cite{yazdanshenas2019costs} is reverse engineering the bitstream of a target FPGA and configuring some routing multiplexers to observe shell wires that pass through the adversary role.
The \emph{FPGAhammer} attack \cite{krautter2018fpgahammer} induces timing faults in the victim logic by repetitive activation patterns and then performs a differential fault analysis to recover an AES key.

\subsubsection*{\textbf{Denial-of-Service Attacks}}
These attacks target crashing an FPGA device by drawing a substantial amount of current either by forming short circuit connections \cite{beckhoff2010short} or by power hammering the FPGA using conventional or camouflaged ring oscillators (ROs) \cite{giechaskiel2019measuring}.
It is also shown in \cite{gnad2017voltage} that excessive voltage drops due to high signal activity in a fraction of the FPGA can result in crashing the entire FPGA.
The authors of \cite{matas2020invited} introduce the GoAhead tool for optimizing and tuning ROs for maximum speed and power consumption.
The optimized RO grid designed by their tool is able to crash a Xilinx Alveo using only 15\% of the FPGA's resources.
The same authors also show that glitch amplifying circuits consisting of XOR trees and long wires can crash a Xilinx Ultrascale+ FPGA using only 0.8\% of the LUTs and 25\% of the routing resources on the device \cite{mataspower}. 
In addition, they implement an FPGA virus-scanning tool, FPGADefender, that detects malicious circuits in a given bitstream to mitigate such attacks.

\subsubsection*{\textbf{Integrity Attacks}}
This class of attacks aims to compromise the victim circuit resulting in incorrect functionality.
A recent work \cite{provelengioscharacterizing} evaluates the ability of malicious circuits to induce delay faults in a neighboring circuit on an FPGA.
They focus mainly on characterizing the response of the FPGA's PDN.
They also study how the effect on the victim circuit changes by changing the time and strength of the voltage drops and the distance between the victim and attacker circuits. 
Using asynchronous RO attacker circuits, they managed to induce timing faults in simple adder circuits that are 42 columns away from the center of the attacker.
In addition to that, they also implement a network of light-weight voltage sensors to monitor voltage gradients and mitigate a potential attack.

In contrast to our work, which falls under the same category, \cite{provelengioscharacterizing} uses asynchronous ROs as their attacker circuit, which can be detected by scanning the given bitstream for combinational loops as in Amazon AWS \cite{rosamazon}.
However, we also demonstrate the feasibility of a similar attack using a vendor-supplied clock gating IP core that can circumvent conventional bitstream checkers.
In addition, they evaluate the integrity attack on a relatively old and small Intel Cyclone V FPGA, while we demonstrate a similar attack on the largest state-of-the-art Intel Stratix 10 device.
Finally, none of the previous studies evaluated such attacks in the context of DL accelerators on FPGAs.
The error resilience of DL models is the main reason for adopting different optimizations such as model pruning and weight quantization with no loss in accuracy \cite{han2015deep, mishra2017wrpn}.
Hence, we investigate the resilience of deployed DL models against timing faults injected by an adversarial FPGA tenant.

\subsection{Hardware-oriented DL Security Vulnerabilities}

There is also previous work on hardware-oriented attacks against DL models.
A commodity DRAM hardware fault, \emph{Rowhammer}, along with memory deduplication (an operating system optimization technique to reduce memory usage) are exploited in \cite{hong2019terminal} to conduct an integrity attack.
It is shown that inducing bit flips in memory causes severe accuracy degradation in many neural network models.
%Surprisingly, across 19 different models, the results show that the ratio of parameters in which a single bit flip causes more than 10\% relative accuracy drop ranges from 40\% to 99\% of the model parameters.
Surprisingly, across 19 different models, 50\% of the model parameters on average are vulnerable to single-bit flips that cause an accuracy drop of more than 10\%.
Another example for an extraction attack is \cite{batina2019csi} in which the authors successfully extract different multi-layer perceptron and convolutional neural network (CNN) models by measuring timing and electromagnetic emanations from two different microprocessors.  
The authors show a methodology for reverse engineering the number of layers, the number of neurons in each layer, the type of activation functions, and the values of model parameters.
A recent denial-of-service attack is demonstrated in \cite{shumailov2020sponge} by crafting malicious DL model inputs that increase energy consumption to drain batteries in IoT devices, or increase the model decision latency to compromise real-time systems.
Similarly to our work, none of these attacks requires physical access to the hardware or prior knowledge of the  DL models (i.e. \emph{black-box} attacks).
While we focus on FPGAs, the first two attacks are carried out on CPUs and the latter is performed on a CPU, a GPU and an ASIC simulator.
The authors of \cite{hua2018reverse} demonstrate an extraction attack against an FPGA-accelerated CNN.
By monitoring the memory traffic through physical access or another side-channel, this attack can extract information about the layer structure and weights.
In contrast, our work focuses on integrity attacks and does not require physical access to the hardware.
%To the best of our knowledge, this work is the first attempt to evaluate DL integrity attacks on FPGAs.
In \cite{rakin2020deep}, the authors use voltage attacks in a multi-tenant FPGA to corrupt the model parameters when transferring them from off-chip memory to on-chip buffers.
This approach would not be effective against persistent-style accelerators that are commonly used in network-connected cloud FPGAs \cite{caulfield2016cloud, hall2020hpipe}.
They also use asynchronous RO attacker circuits which can be easily filtered by conventional bitstream checkers.
Our work does not assume the need for transferring weights from off-chip memory and uses combinational-loop-free attacker circuits that cannot be detected by scanning the bitstream.

%However the first two attacks were carried out on CPUs and the latter was performed on a CPU, a GPU and an ASIC simulator; to the best of our knowledge, this is the first attempt to evaluate DL integrity attacks on FPGAs.
\section{Attacker Circuits}

In this section, we first introduce the preliminaries of timing violations, PDNs and voltage attacks on FPGAs that are necessary to understand the rest of the paper.
Then, we introduce the attacker circuits used in our experiments.

\subsection{Timing Violations}
For correct operation, the clock period of a digital circuit must ensure the delay between two clock edges is larger than the time taken by a signal to travel the longest combinational path between two flip-flops (FFs).
We can formulate this relation using the following inequality:
\begin{equation}
    t_{clk} \geq t_{c2q} + t_{comb} + t_{setup} + t_{skew}
    \label{eq:timing}
\end{equation}
where $t_{clk}$ is the clock period, $t_{c2q}$ is the the clock-to-output delay of the source FF, $t_{comb}$ is the longest combinational path delay from source to destination, $t_{setup}$ is the time the input to the destination FF needs to be stable to be captured correctly, and $t_{skew}$ is the largest difference in clock delay between the source and destination FFs, including variability and jitter effects.
A \emph{setup timing violation} happens if inequality (\ref{eq:timing}) does not hold for any path in a given circuit, and can result in a faulty output being captured by the destination FF.
For the rest of the paper, we simply refer to these violations as \emph{timing violations}, as we are not concerned with hold timing violations in this work.

\begin{figure}[t!]
\centering
  \includegraphics[width=\linewidth]{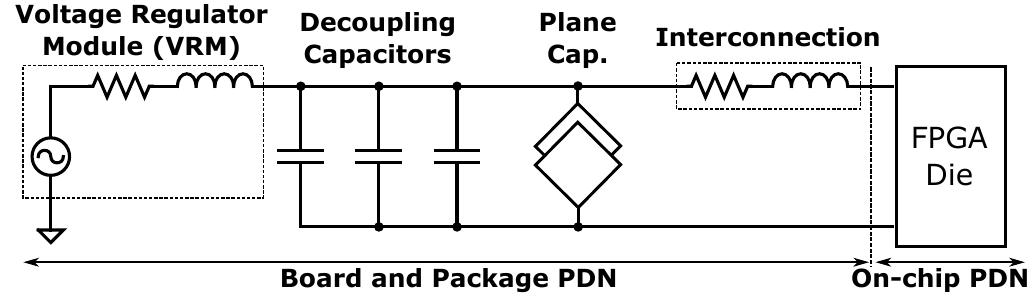}
  \vspace{-0.1cm}
  \caption{Simplified diagram of the PDN topology.}
  \label{fig:pdn}
  \vspace{-0.5cm}
\end{figure}

\subsection{Attack Methodology}
A simplified schematic showing the main components of the FPGA's PDN is shown in Fig. \ref{fig:pdn}.
An on-board voltage regulator module (VRM) is used to convert the board voltage level down to the die voltage level.
After that, the on-board decoupling capacitors and the parallel plane capacitance between the board power and ground layers help filter out the voltage noise and respond to rapid variations in load \cite{intel2015pdn}.
Finally, the connection to the FPGA die is made through the ball grid array based package, and creates a series-connected resistance and inductance.
The on-chip FPGA PDN itself can be viewed as distributed RC network or even simplified as an equivalent RC circuit\cite{klokotov2014distributed}.
Consequently, the voltage drop between the output of the VRM and any end point on the FPGA fabric can be captured by equation (\ref{eq:voltage_drop}).

\vspace{-0.3cm}
\begin{equation}
    %V_{drop} = iR + L\frac{di}{dt}
    %V_{drop} = I(s)R + I(s)Z_{AC}(s)
    V_{drop}(s) = I(s)Z_{PDN}(s)
    \label{eq:voltage_drop}
\end{equation}
where $Z_{PDN}(s)$ is the frequency domain impedance resulting from the combination of board, package and chip resistance and inductance along with the various decoupling capacitances shown in Fig. \ref{fig:pdn}. 
For steady-state current, only the resistive component of $Z_{PDN}$ is relevant and gives rise to a \emph{steady state} drop, i.e. $IR$. 
Changes in the load current give rise to a \emph{transient drop} due to the inductive component of $Z_{PDN}$, i.e. $L\frac{di}{dt}$. 
This drop is partially mitigated by the different capacitive components in the PDN that can supply current for short times (i.e. at high frequencies).
The results in \cite{klokotov2014distributed} also show that the $V_{drop}$ experienced at different locations on the FPGA die can be different as a result of different spatial distances from the decoupling capacitors (i.e. as the distance increases, the decoupling capacitors are less effective in mitigating voltage fluctuations).
Therefore, an effective attacker circuit would be one that draws a large amount of current, resulting in a large $IR$ drop, and at the same time experiences abrupt changes in the amount of drawn current, resulting in a large transient drop. 
As a result of the voltage drop created by the attacker circuit, the signal propagation time through the combinational path ($t_{comb}$) increases, breaking inequality (\ref{eq:timing}) and inducing timing violations in a neighboring victim circuit on the FPGA.

\vspace{-0.1cm}
\subsection{Attacker Circuit 1: Asynchronous ROs}
\begin{figure}
    \centering
    
    \begin{subfigure}[b]{0.5\linewidth}
        \centering
        \includegraphics[width=0.5\textwidth]{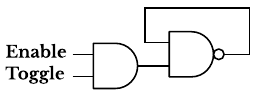}
        \caption{Asynchronous ROs.}
        
        \includegraphics[width=\textwidth]{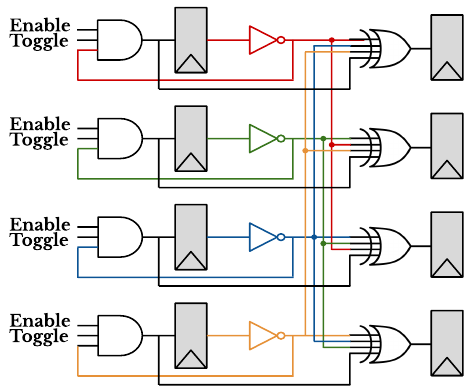}
        \caption{Clock-gated garbled XORs.}
    \end{subfigure}~
    \begin{subfigure}[b]{0.45\linewidth}
    \centering
    \includegraphics[width=\textwidth]{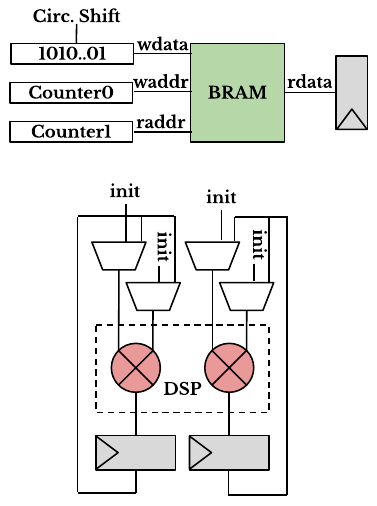}
    \caption{Clock-gated hybrid logic.}
    \end{subfigure}
    
    \caption{Attacker circuits.}
    \label{fig:attackers}
\end{figure}

An asynchronous RO can be implemented as a NAND gate whose output is fed back into its input as in Fig. \ref{fig:attackers}a.
When the enable and toggle signals are set to high, the output of the NAND gate oscillates at a frequency determined by the interconnect delay between the gate's output and input.
For a Stratix 10 FPGA, this delay is in the order of hundreds of picoseconds, resulting in a very high oscillation frequency on the order of several GHz.
A large number of these ROs implemented in the FPGA's soft fabric is capable of drawing a large amount of current due to their very fast switching activity.
In addition, when the toggle signal is switched periodically with a specific frequency and duty cycle, it results in an abrupt change in the amount of drawn current. 
For our adversary role, we implement 20 RO grids, similar to that used in \cite{krautter2018fpgahammer}, such that each grid is a 32$\times$1024 array of ROs.
Each grid also has a set of three 32-bit control registers responsible for enabling/disabling parts of the grid, as well as controlling the frequency and duty cycle of the toggle signal.

Asynchronous ROs are typically used in most of the prior work to carry out voltage attacks (e.g. \cite{krautter2018fpgahammer,provelengioscharacterizing,matas2020invited}) however, it is not a practical adversarial circuit since bitstream checkers can detect such combinational loops and automatically reject the malicious bitstream.
This is a defense technique that is well-studied in academia \cite{matas2020invited} and also commonly deployed in Amazon AWS \cite{rosamazon}.
Therefore, we introduce the following two attacker circuits to circumvent this defense technique.

%\begin{figure}[t!]
%  \centering
%  \includegraphics[width=0.95\linewidth]{Figures/attackers.pdf}
%  \vspace{-0.3cm}
%  \caption{Attacker circuits: (a) asynchronous ROs, (b) clock-gated garbled XORs, and (c) clock-gated hybrid toggling logic.}
%  \label{fig:attackers}
%  \vspace{-0.2cm}
%\end{figure}

\begin{table}[t!]
    \centering
    \caption{Resources dedicated to victim and attacker roles and resource utilization for each attacker circuit (CG: clock-gated)}
    \vspace{-0.2cm}
    \begin{tabular}{C{1.4cm} C{2.2cm} C{1.8cm} C{1.8cm}} 
    \hline\\ [-1.8ex]
    \multicolumn{4}{c}{\textbf{Resources dedicated to each role}}\\
    \textbf{Role} & \textbf{ALMs} & \textbf{DSPs} & \textbf{BRAMs} \\ [0.25ex] 
    \hline\\ [-1.8ex]
    Victim    & 395,040 (43\%) & 2,488 (44\%) & 5,142 (44\%)\\ 
    Adversary & 406,560 (44\%) & 2,632 (46\%) & 5,407 (47\%)\\ [0.25ex]
    \hline\\ [-1.8ex]
    \multicolumn{4}{c}{\textbf{Resources utilized by each attacker circuit}}\\
    \textbf{Attacker} & \textbf{ALMs} & \textbf{DSPs} & \textbf{BRAMs} \\ [0.25ex]
    \hline\\ [-1.8ex]
    ROs       & 330,220 (36\%) & 0 (0\%) & 0 (0\%) \\
    CG XORs   & 344,064 (37\%) & 0 (0\%) & 0 (0\%) \\
    CG Hybrid & 304,589 (33\%) & 2,400 (42\%) & 4,768 (41\%) \\ [0.25ex]
    \hline
    \end{tabular}
    \label{tab:resources}
    \vspace{-0.3cm}
\end{table}

\vspace{-0.1cm}
\subsection{Attacker Circuit 2: Clock-gated Garbled XORs}
The basic building block of the second attacker circuit (shown in Fig. \ref{fig:attackers}b) was previously used in \cite{shen2019fast} to study transient effects on FPGAs, but was never evaluated in the context of voltage attacks before.
It consists of four 4-input XOR gates, whose inputs are chosen from four toggle registers and one delayed output. 
The toggle registers change state each cycle and the output of the XOR gates change if any input changes, so this results in a high switching activity. 
For our adversary role, we implement six attacker grids of size 32,768 garbled XOR blocks each.
We clock the whole adversary role using a fast 750 MHz clock, as we are not concerned with meeting timing in the attacker circuit anyway.
As mentioned in the Stratix 10 power management user guide, clock gating a large portion of the FPGA could cause significant current change over a short time period when the gated circuitry is enabled or disabled \cite{s10power}. %which can potentially result in a large $L\frac{di}{dt}$ voltage drop as discussed in the previous section. 
Enabling a clock suddenly changes the toggle rates of all the gated FFs.
In addition, the clock networks in modern FPGAs are composed of large wires in the higher metal layers, and therefore have high capacitance. 
Hence, large transient events can happen when the clock is started again after gating. 
These transient effects have very sharp edges (i.e. high-frequency components) which have to be handled by the more limited on-die decoupling capacitances. 
Therefore, if the transient events are strong enough (beyond what the on-die capacitances can fully handle), they can cause voltage drops and thus induce timing faults in the victim circuit.
We use the Intel clock control IP core \cite{s10clk} to perform root clock gating for the attacker circuits in the adversary role.
The clock gating signal is toggled periodically at a frequency and duty cycle specified using control registers. 
Unlike asynchronous ROs, this attacker circuit uses a vendor-supplied IP core for clock gating and does not contain any combinational loops. 
Thus, it poses as an unsuspicious circuit and would pass any conventional bitstream checker.

%\begin{figure}[t!]
%  \centering
%  \includegraphics[width=0.8\linewidth]{Figures/victim_system.pdf}
%  \vspace{-0.4cm}
%  \caption{(a) Timing violation detection circuit, and (b) overall system block diagram used for the integrity attack characterization (CL: chain length).}
%  \label{fig:victim}
%  \vspace{-0.3cm}
%\end{figure}

\begin{figure}
    \centering
    
    \begin{subfigure}[b]{0.75\linewidth}
        \centering
        \includegraphics[width=\textwidth]{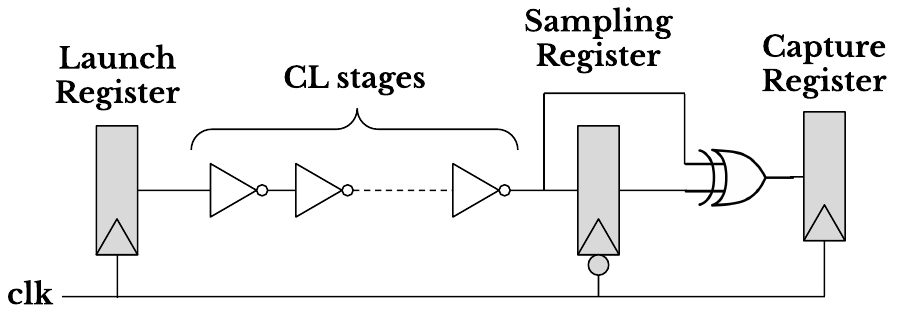}
        \caption{}
    \end{subfigure}
    
    \begin{subfigure}[b]{0.75\linewidth}
    \centering
    \includegraphics[width=\textwidth]{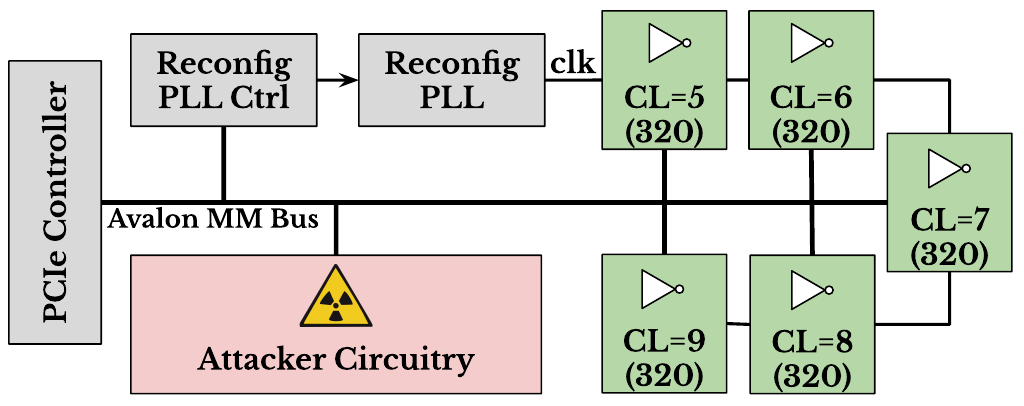}
    \caption{}
    \end{subfigure}
    
    \caption{(a) Timing violation detection circuit, and (b) overall system used for the integrity attack characterization (CL: chain length).}
    \label{fig:victim}
\end{figure}

\begin{figure}[t!]
  \centering
  \includegraphics[width=0.7\linewidth]{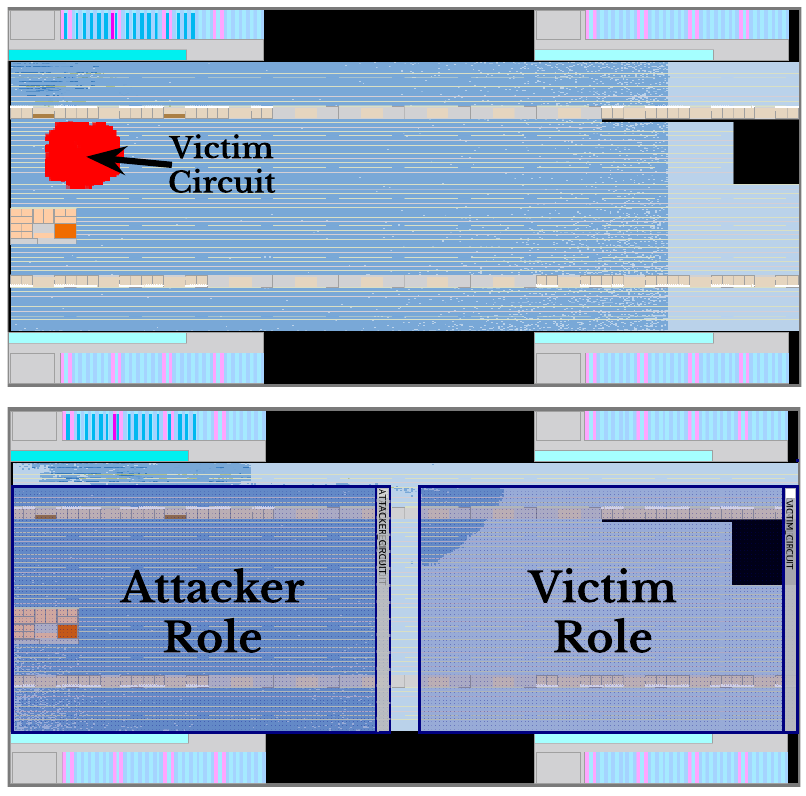}
  \caption{Chip planner view for the non-floorplanned (upper) and floorplanned (bottom) implementations of the FPGA design with the clock-gated garbled XOR attacker. In the non-floorplanned implementation, the CAD tool decides to place the attacker circuit around the victim arrays. In the floorplanned one, the same attacker circuit is more densely-packed in an isolated role.}
  \label{fig:chip_planner}
  \vspace{-0.5cm}
\end{figure}

\begin{figure*}[t!]
  \centering
  \includegraphics[width=1\linewidth]{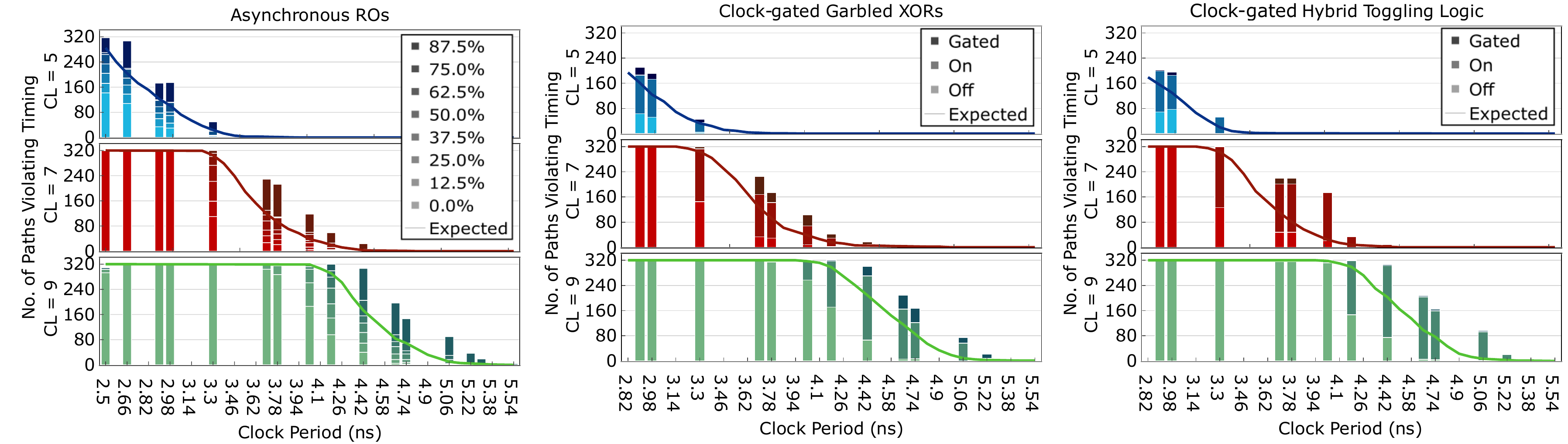}
  \vspace{-0.3cm}
  \caption{The number of timing violation detection paths with flagged faults for the floorplanned implementations of the asynchronous RO (left), clock-gated garbled XOR (middle), and clock-gated hybrid toggling logic (right) designs. The bars represent the number of paths violating timing measured on hardware at different operating clock periods, compared to the lines that represent the numbers of paths expected to violate timing based on Quartus timing reports. At points where the bar value is higher than the line, the attacker circuit successfully induced timing faults in paths that Quartus reported were safe.}
  \label{fig:bars}
  \vspace{-0.3cm}
\end{figure*}

\subsection{Attacker Circuit 3: Clock-gated Hybrid Toggling Logic}
The first two attacker circuits are implemented exclusively using the soft logic in the adversary role, leaving all block memories (BRAMs) and digital signal processing (DSP) blocks not utilized.
The third attacker circuit aims to maximize the switching activity in the adversary role by using all the DSP blocks and BRAMs as shown in Fig. \ref{fig:attackers}c.
During the attack, alternating chessboard patterns are sequentially written to every address of all BRAMs, each multiplier in all the DSP blocks is seeded with random initial inputs that are multiplied and fed-back as inputs to the DSP block, and the remainder of the soft-logic is used to implement garbled XOR blocks, shown in Fig. \ref{fig:attackers}b.
Similarly to the garbled XORs attacker, this circuit is also clock gated periodically and does not contain any suspicious logic that would be flagged by a typical bitstream checker.

Table \ref{tab:resources} lists the resources dedicated (available) to each of the victim and attacker roles.
We assume only two roles per device and leave out $\sim$13\% of the device to be used for the FPGA shell implementation.
The table also summarizes the resource utilization of the attacker role using each of the three attacker circuits described in this section.
\section{Attack Characterization}
\label{sec:attack_char}

\subsection{Timing Violation Detection Circuit}
In order to demonstrate the ability of the attacker circuits to induce timing violations in a given logically and physically isolated victim circuit, we implement a simple timing violation detection circuit, similar to that used in \cite{gojman2014grok}, as illustrated in Fig. \ref{fig:victim}a.
In this circuit, a signal is launched from the \emph{launch register} at the rising clock edge, passes through a set of $CL$ inverter stages, and is finally captured by the \emph{sampling register} at the following falling clock edge.
We use Quartus synthesis pragmas to implement the $CL$ inverters in $CL$ distinct look-up tables and prevent the compiler from reducing them into a single equivalent logic element.
At the next rising clock edge, the value of the sampling register and the output of the last inverter stage are compared, and the \emph{capture register} is set to high if they are different.
The value of the capture register is then latched such that if it is once set to high, it stays high until the end of the experiment.
Under normal operating conditions, if the propagation delay through the $CL$ inverter stages is less than half the clock period, the output of the last inverter will always be the same as that of the sampling register.
Thus, the capture register will never be set to high.
However, if the adversary causes a voltage drop significant enough to increase the inverter chain propagation delay beyond half a clock period, a timing violation is flagged by setting the capture register high.

\subsection{Experimental Setup}
Our experimental setup consists of a Terasic DE10-Pro board \cite{terasic} with the largest monolithic state-of-the-art Intel Stratix 10 2800 FPGA attached as a PCIe accelerator card to an Intel Xeon E5-2650 server with 12 double-threaded cores and 94 GBs of RAM.
All the FPGA designs used in our experiments are synthesized, placed and routed using Intel Quartus Prime Pro 19.3.
Fig. \ref{fig:victim}b shows a block diagram of the system we implement on the FPGA.
The victim role is occupied by five arrays, each of which has 320 of the timing violation detection chains in Fig. \ref{fig:victim}a.
The five arrays have five different chain lengths (i.e. $CL$ values) ranging from 5 up to 9 inverter stages.
The safe operating frequencies for the timing violation detection arrays (i.e. frequencies in which the capture registers are never set to high), as reported by Quartus at the slow 100$^{\circ}$C corner and averaged over 6 compilations, are listed in Table \ref{tab:freq}. 
All the victim arrays, as well as the attacker circuits, are wrapped as IP cores with an Avalon memory-mapped slave interface \cite{avalon} and connected to the PCIe controller over an Avalon bus as illustrated in Fig. \ref{fig:victim}b.
In addition, we instantiate a reconfigurable phase-locked loop (PLL) to generate the clock for all victim arrays.
We implement a PLL controller that adjusts the frequency of the clock generated by the reconfigurable PLL from 150 MHz to 400 MHz based on host commands sent over PCIe. 

\begin{table}[t!]
    \centering
    \caption{Safe operating frequency for timing violation detection chain arrays with different $CL$ values at the slow 100$^\circ$C corner.}
    \vspace{-0.25cm}
    \begin{tabular}{C{2.4cm} C{0.7cm} C{0.7cm} C{0.7cm} C{0.7cm} C{0.7cm} C{0.7cm}} 
    \hline\\ [-1.8ex]
    \textbf{Chain Length (CL)} & 5 & 6 & 7 & 8 & 9 \\ [0.25ex] 
    \hline\\ [-1.8ex]
    \textbf{Safe Freq. (MHz)} & 260.3 & 240.6 & 211.4 & 198.5 & 182.7 \\ [0.25ex] 
    \hline\\
    \end{tabular}
    \label{tab:freq}
    \vspace{-0.6cm}
\end{table}

%\begin{figure*}[t!]
%  \centering
%  \includegraphics[width=0.95\linewidth]{Figures/heatmaps.pdf}
%  \vspace{-0.3cm}
%  \caption{Heatmaps illustrating the number of paths violating timing for: (a) the non-floorplanned and floorplanned implementations of the asynchronous RO attacker design, and (b) the floorplanned implementations of the clock-gated garbled XOR (top) and clock-gated hybrid toggling logic (bottom) attacker designs. Black dashed lines represent the maximum safe operating frequency reported by Quartus.}
%  \label{fig:heatmaps}
%  \vspace{-0.4cm}
%\end{figure*}

\begin{figure*}
    \centering
    
    \begin{subfigure}[b]{0.62\linewidth}
        \centering
        \includegraphics[width=\textwidth]{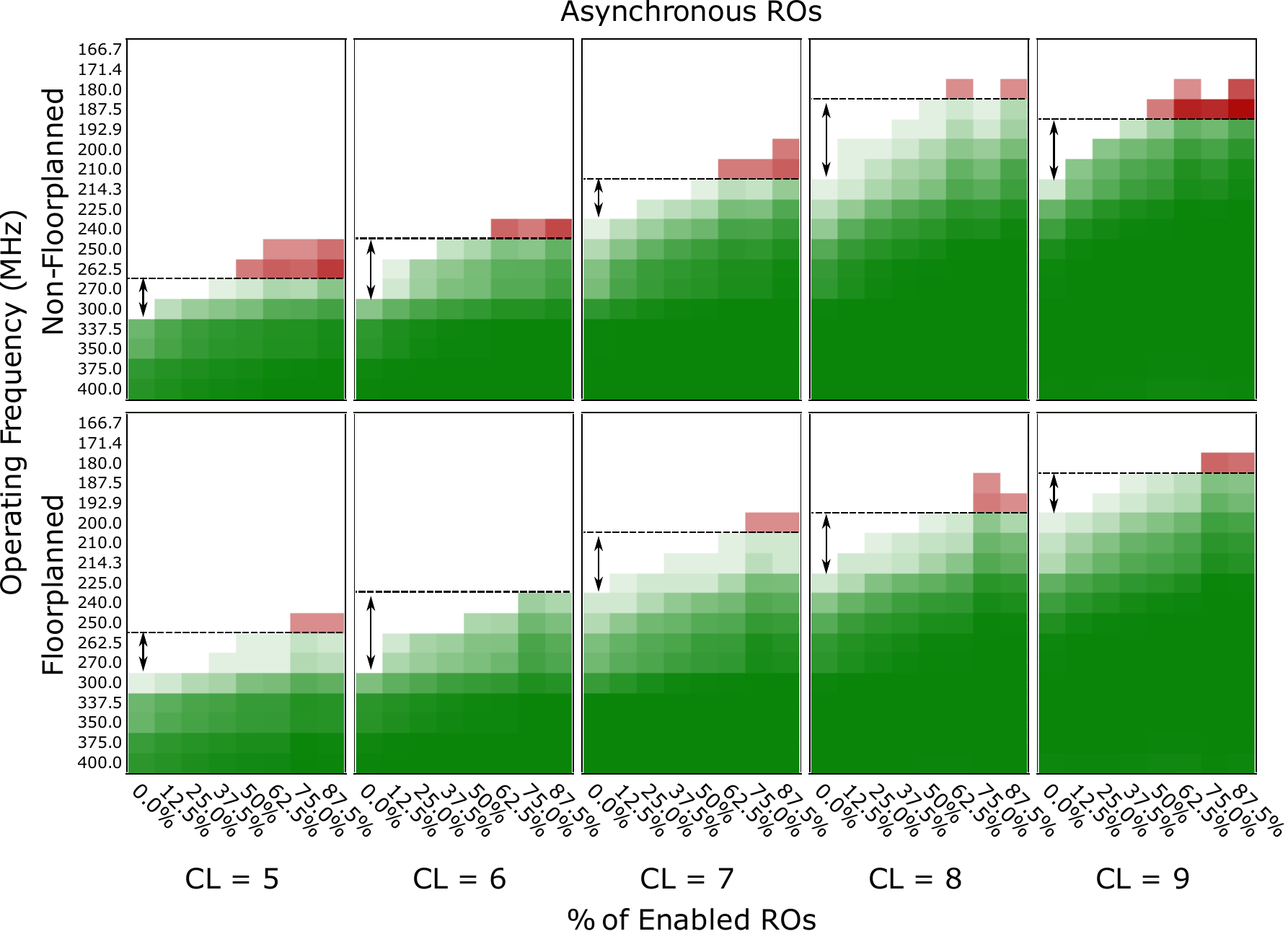}
        \caption{}
    \end{subfigure}~
    \begin{subfigure}[b]{0.36\linewidth}
    \centering
    \includegraphics[width=\textwidth]{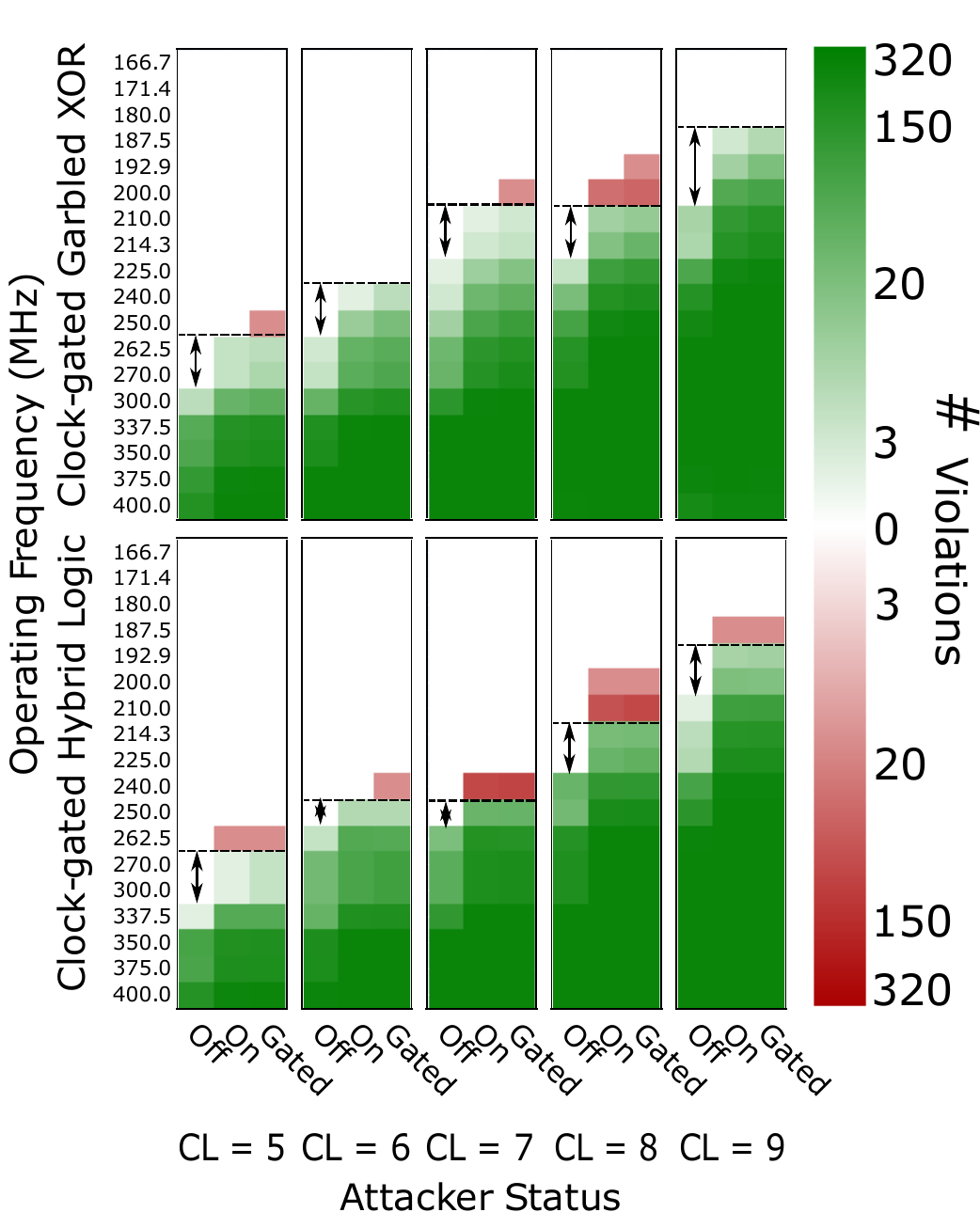}
    \caption{}
    \end{subfigure}
    
    \caption{Heatmaps illustrating the number of paths violating timing for: (a) the non-floorplanned and floorplanned implementations of the asynchronous RO attacker design, and (b) the floorplanned implementations of the clock-gated garbled XOR (top) and clock-gated hybrid toggling logic (bottom) attacker designs. Black dashed lines represent the maximum safe operating frequency reported by Quartus.}
    \label{fig:heatmaps}
    \vspace{-0.4cm}
\end{figure*}

For each of our attacker circuits, we generate two different bitstreams; a non-floorplanned implementation in which the CAD tool decides how to place the victim and attacker circuits, and a floorplanned one to isolate the victim and attacker circuits in two different roles emulating a realistic multi-tenant FPGA.
The chip planner view of both implementations for the clock-gated garbled XORs attacker is shown in Fig. \ref{fig:chip_planner}.
For the non-floorplanned implementation, the tool decides to place the attacker circuit surrounding the victim arrays.
In the floorplanned implementation, each role is defined as a logic locked partition with routing constrained to the partition area to mimic a dynamically reconfigurable partition in a virtualized FPGA.

A C++ program running on the host CPU configures the victim arrays' operating frequency and the attackers' control registers via PCIe.
Then, it enables the attacker circuits for $250\times10^6$ cycles, before disabling them and then reading back the capture registers of the timing violation detection circuits to determine the number of faults induced by the attack.

\vspace{-0.2cm}
\subsection{Characterization Results}

\subsubsection*{\textbf{Attack Feasibility}}
The line graphs in Fig. \ref{fig:bars} show the number of paths expected to violate timing in the victim circuit at different operating frequencies based on the timing report produced by Quartus.
The graphs are for violation detection circuits with $CL$ values 5, 7 and 9 in the floorplanned implementations. 
However, other $CL$ values and the non-floorplanned implementations show similar trends and thus we  omit them for brevity.
For clock periods larger than the longest path delay, none of the paths is expected to violate timing.
As the clock period decreases, more paths start to violate timing until all of them have path delays that are larger than the operating clock period.

In the same figures, these expected values are compared to the stacked bars reporting the actual number of timing violations measured on hardware at different frequencies of the clock produced by the reconfigurable PLL.
When the attacker circuit is disabled (indicated by 0\% ROs or off in Fig. \ref{fig:bars}), the numbers of measured violations (bars) are always lower, and in some cases by a large gap, than that reported by Quartus (line). 
This establishes that Quartus' timing analysis introduces a considerable safety margin to account for aging effects, process variations, etc.
For the asynchronous RO attacker, the number of measured violations increases as we increase the percentage of enabled ROs in the adversary roles.
At every point the bar value is higher than the line, it means that the attacker circuit managed to induce timing faults in paths that Quartus reported were safe (i.e. have lower path delay than the operating clock period).
Enabling more than 87.5\% of the ROs in the adversary role, which is equivalent to $\sim$38\% of the FPGA resources, crashes the entire FPGA board and requires power-cycling the whole server before re-programming the FPGA.
As discussed in Section \ref{sec:background}, this is considered a denial-of-service attack. 
Since our main focus is on integrity attacks, we are most interested in attacker circuits that draw as much current as possible to increase the induced timing violations, but still do not result in crashing the entire board.
%However, we are not interested in this type of attack in this work.
For the clock-gated garbled XORs and hybrid toggling logic attackers, the number of paths violating timing increases significantly by up to $\sim$200 additional failing paths when the attacker circuit is enabled without periodic clock gating (On), and increases further ($\sim$45 more paths) as a result of periodic clock gating (Gated). 
The results show that, in many cases, the 3 attacker circuits were able to cause a higher number of paths to violate timing than Quartus reported, and thus demonstrate the feasibility of our integrity attack.

\subsubsection*{\textbf{Floorplanning Effect}}
Fig. \ref{fig:heatmaps}a presents a heatmap of the number of paths violating timing at various operating frequencies (vertical axis) and $CL$ values for both the floorplanned (top) and non-floorplanned (bottom) implementations with various percentages of enabled ROs (horizontal axis). 
Timing faults induced at frequencies higher than the maximum operating frequencies reported by Quartus (i.e. expected violations) are colored in green, while those induced at presumably safe frequencies are colored in red.
The latter ones are the most interesting for our study as they represent successful integrity attacks at operating conditions that are supposed to be safe according to Quartus' timing analysis. 

Fig. \ref{fig:heatmaps}a shows that we can induce timing faults at frequencies Quartus considers safe with approximately 50\% of ROs active when the roles are not floorplanned, and with about 75\% of ROs active when the roles are physically isolated.
The results show that physically isolating the victim and the attacker circuits in two distinct roles can help mitigate the severity of the attack and considerably reduce the number of induced timing violations.
However, it is not sufficient to entirely protect the victim circuit from its malicious neighboring roles.
The comparison between the non-floorplanned and floorplanned implementations of the systems with the other two attacker circuits also corroborates the same conclusion; however, we leave them out for brevity.
This observation advocates for the potential need for redesigning the on-chip PDN with multiple separate voltage islands in future FPGA architectures to enable secure multi-tenant FPGA virtualization.
In this case, voltage disruptions caused by an adversary would not affect a victim circuit that belongs to a different voltage island.

\subsubsection*{\textbf{Non-RO-based Attacker Circuits}}
In Fig. \ref{fig:heatmaps}b, we show the same heatmaps for the floorplanned implementation of the systems with the clock-gated garbled XORs and clock-gated hybrid toggling logic attackers.
The results show that the clock-gated garbled XOR attacker circuit can be almost as effective as the asynchronous ROs attacker circuit at its highest capacity (i.e. with 87.5\% of the ROs enabled).
It also has the additional advantage of being undetectable by conventional bitstream checkers as a potentially malicious circuit.
Moreover, clock-gating is an increasingly popular power saving technique in FPGA designs, raising the possibility that even a benevolent circuit could unintentionally inject timing faults in itself or in neighboring roles by aggressively applying clock gating on a power-hungry computation.
In addition, as shown in Fig. \ref{fig:heatmaps}b, utilizing the hard blocks (i.e. BRAMs and DSPs) in the attacker role to create higher switching activity can further enhance the severity of the attack compared to the garbled XORs attacker that only utilizes soft logic resources.

\subsubsection*{\textbf{Quartus Timing Safety Margins}}
We can also quantify the safety margins added by the Quartus timing analyzer by measuring the difference between the maximum safe operating frequency reported by Quartus and the frequency at which we witness the first timing violations with the attacker circuit disabled.
These margins are highlighted with the black arrows (safety margin) and dashed lines (Quartus reported maximum frequency) in Fig. \ref{fig:heatmaps}a and \ref{fig:heatmaps}b. 
Our hardware measurements show that, on average across multiple compilations of our test designs with different attacker circuits and floorplanning setups, Quartus adds to the reported clock period a safety margin of 13\% of the longest path delay, with a maximum and minimum of 21\% and 5\%, respectively.
Our experiments are conducted on a new Stratix 10 card.
Over time, we would expect this measured safety margin to be reduced by transistor aging effects, further escalating the severity of voltage attacks. 

%\begin{figure*}[t!]
%  \centering
%  \includegraphics[width=\linewidth]{Figures/hpipe.pdf}
%  \vspace{-0.4cm}
%  \caption{Effect of induced timing faults on the prediction accuracy of the MobileNet-V1 CNN running on the victim HPIPE at different operating frequencies with the: (a) clock-gated garbled XORs and (b) clock-gated hybrid toggling logic attackers. The bars represent the value of the angle between the prediction vector at each corresponding frequency compared to the safe operating frequency 540 MHz.}
%  \label{fig:hpipe}
%  \vspace{-0.5cm}
%\end{figure*}

\begin{figure*}
    \centering
    
    \begin{subfigure}[b]{0.47\linewidth}
        \centering
        \includegraphics[width=\textwidth]{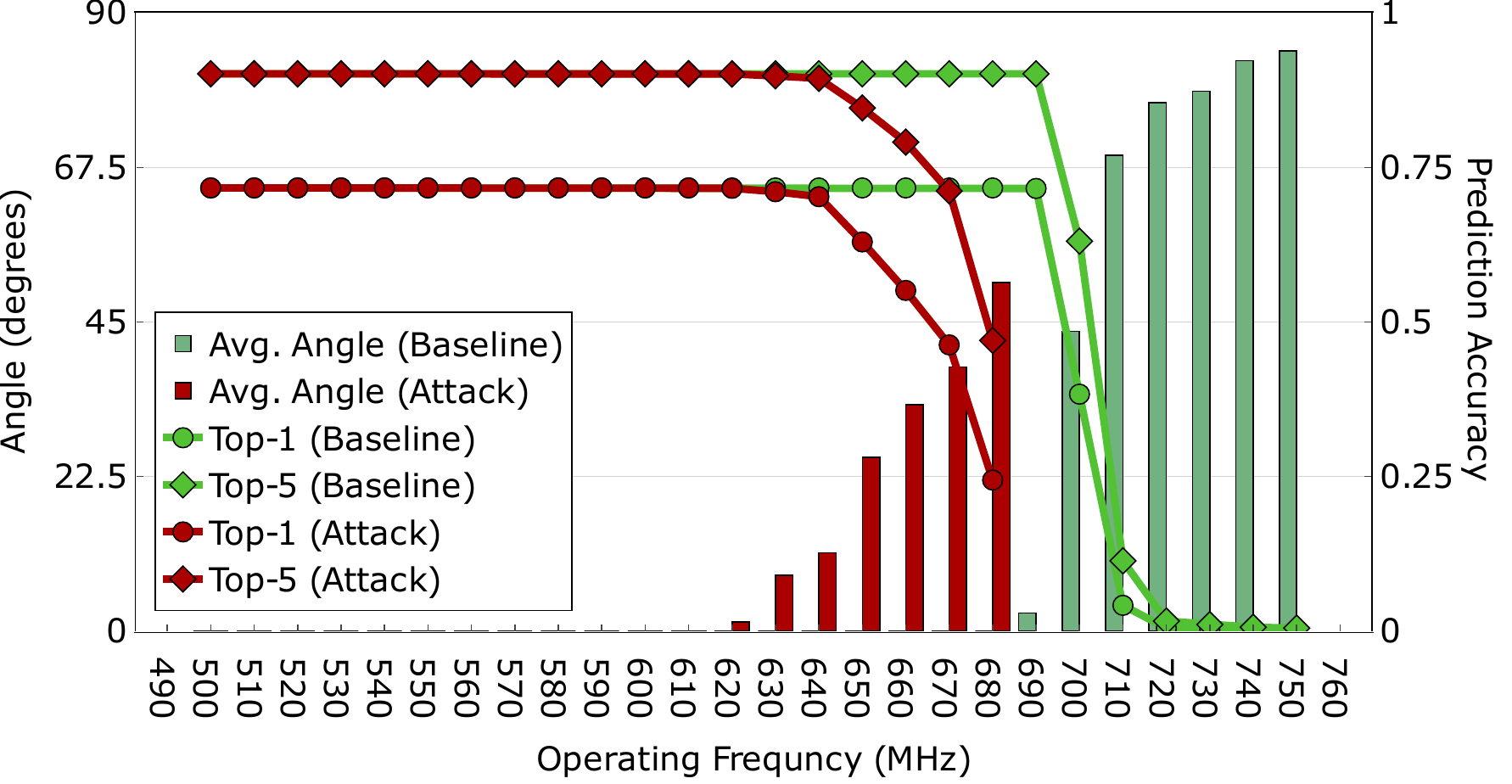}
        \caption{}
    \end{subfigure}~
    \begin{subfigure}[b]{0.47\linewidth}
    \centering
    \includegraphics[width=\textwidth]{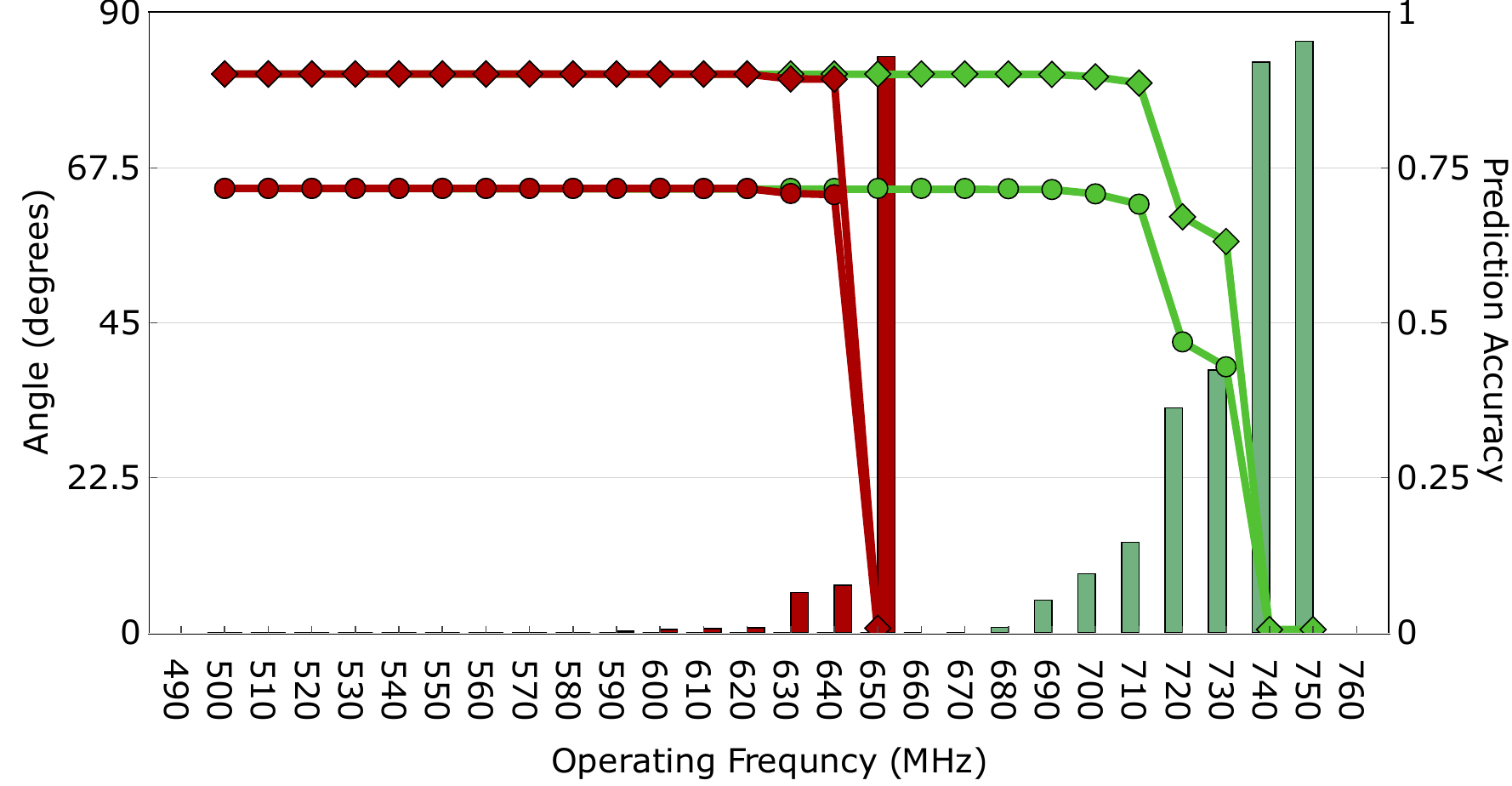}
    \caption{}
    \end{subfigure}
    \vspace{-0.2cm}
    %\caption{Heatmaps illustrating the number of paths violating timing for: (a) the non-floorplanned and floorplanned implementations of the asynchronous RO attacker design, and (b) the floorplanned implementations of the clock-gated garbled XOR (top) and clock-gated hybrid toggling logic (bottom) attacker designs. Black dashed lines represent the maximum safe operating frequency reported by Quartus.}
    \caption{The effect of voltage attacks on the victim HPIPE accelerator in a physically isolated role on a multi-tenant FPGA when using: (a) clock-gated garbled XORs, and (b) clock-gated hybrid toggling logic attackers. The lines represent top-1 and top-5 prediction accuracy and the bars represent the angle between the prediction vector at a given operating frequency and that at the 540 MHz safe frequency. The green and red traces show results with attacker circuits disabled and enabled, respectively.}
    \label{fig:hpipe}
    \vspace{-0.4cm}
\end{figure*}

\section{Attacking a DL Accelerator}

DL models are known for their error resilience, which enabled the adoption of several optimization techniques for efficient DL computation such as model compression \cite{han2015deep} and weight quantization \cite{mishra2017wrpn}.
Studies have shown that $\sim$90\% of the model weights can be pruned without affecting the model accuracy \cite{han2015learning}.
In addition, a myriad of numerical precisions are used in DL inference, ranging from single-precision floating point to 16-bit fixed point and down to even ternary and binary precisions, often with negligible or no accuracy degradation \cite{boutros2018embracing}.
In this section, we study the resilience of DL models against another source of error: timing faults in the hardware circuit performing the DL model computations.
We measure the effect of voltage drops induced by an adversary FPGA tenant on the prediction accuracy of a DL accelerator running ImageNet classification, and introduce the accelerator's operating frequency as another knob that offers a performance/accuracy tradeoff similar to sparsity and precision.

\vspace{-0.1cm}
\subsection{The Victim DL Accelerator: HPIPE}
For our experiments, we use HPIPE, a state-of-the-art CNN accelerator implemented and optimized specifically for FPGAs. 
It comes with a complete flow that takes a TensorFlow description of the CNN model along with a specification of available FPGA resources, and produces an optimized FPGA-based hardware accelerator for this model given the specified resource constraints. 
The produced accelerator has a cross-layer pipelined spatial architecture that exploits both reduced precision and weight pruning to fit the complete model in on-chip BRAMs.
The HPIPE compiler maximizes the DSP block utilization and produces a highly-pipelined, physically-aware accelerator architecture that can run at very high clock frequencies.
On the largest Intel Stratix 10 FPGA, HPIPE runs ResNet-50 batch-1 inference at almost 4$\times$ the throughput of an Nvidia V100 GPU.
We refer the reader to \cite{hall2020hpipe} for a more detailed description the HPIPE architecture and compiler.

\begin{table}[t!]
    \centering
    \caption{The victim HPIPE resources, performance \& accuracy.}
    \begin{tabular}{C{2.8cm} C{3.5cm}} 
    \hline\\ [-1.8ex]
    \textbf{Model} & MobileNet-V1 (Dense) \\ 
    \textbf{Precision} & 16-bit fixed point \\[0.25ex] 
    \hline\\ [-1.8ex]
    \textbf{ALMs} &  210,698 (23\%)\\ 
    \textbf{DSPs} &  1,501 (26\%)\\
    \textbf{BRAMs} &  5,068 (44\%)\\ [0.25ex] 
    \hline\\ [-1.8ex]
    \textbf{Max. Frequency} & 540 MHz \\
    \textbf{Throughput (batch-1)} & 1,652 images/sec\\
    \textbf{Latency (batch-1)} & 1.99 ms\\
    \textbf{Top-1 Accuracy} & 71.7\%\\
    \textbf{Top-5 Accuracy} & 90.16\%\\[0.25ex] 
    \hline\\
    \end{tabular}
    \label{tab:hpipe}
    \vspace{-0.8cm}
\end{table}

\vspace{-0.1cm}
\subsection{Experimental Setup}
We use the same experimental setup as that explained in Section \ref{sec:attack_char}.
We replace the timing violation detection circuits with an HPIPE instance in the victim role and attack it using our clock-gated garbled XORs and hybrid toggling logic attacker circuits.
Both the HPIPE role and the attacker role are floorplanned to be physically separate, which makes the attack more realistic.
The HPIPE instance we use runs the MobileNet-V1 model and is restricted to the victim role resources in our multi-tenant FPGA setup. 
Table \ref{tab:hpipe} summarizes the FPGA resource utilization results, as well as the operating frequency and baseline performance/accuracy of our victim HPIPE instance. 
The reconfigurable PLL in our test system, shown in Fig. \ref{fig:victim}b, is set to produce clock frequencies ranging from 500-750 MHz.
At each operating frequency during our experiments, we run inference over 50,000 images and record the prediction accuracy with the attacker circuit disabled.
Then, we enable the attacker circuit and run inference over the same 50,000 images again to measure the accuracy under the effect of induced voltage drops.
To capture the numerical differences in the model outputs, we also measure the angle between the prediction vector (i.e. output vector of the last layer before applying softmax) at a given operating frequency and that at the 540 MHz safe frequency, averaged across all 50,000 inputs, using the following formula:

\begin{equation}
    Angle_{avg} = \frac{1}{N} \sum_{i=0}^{N}\frac{P_{s_i}}{||P_{s_i}||} \cdot \frac{P_{f_i}}{||P_{f_i}||}
\end{equation}
where $N$ is the total count of test inputs, $||.||$ is the L2-norm, $P_{s_i}$ and $P_{f_i}$ are the prediction vectors at the safe frequency and the operating frequency of the $i^{th}$ input, respectively.
A non-zero angle value with no degradation in accuracy means that the model is resilient to the timing faults induced in its computations at a given operating frequency.

\subsection{Attack Results}
\subsubsection*{\textbf{Safe Overclocking}}
Fig. \ref{fig:hpipe}a and \ref{fig:hpipe}b show the results of our experiments for the two systems with the clock-gated garbled XORs and hybrid toggling logic attackers, respectively.
With the attacker circuits disabled, the hardware measurements show that we can overclock the DL accelerator to a 27\%-31\% higher clock frequency without affecting the prediction accuracy.
This margin for safe overclocking is considerably bigger than the timing safety margin introduced by the Quartus timing analyzer which we quantified in Section \ref{sec:attack_char}.
The reason for this is the resilience of DL models that, even with induced computational errors, can still predict the correct image classification.
If we increase the clock frequency beyond that, the prediction accuracy degrades rapidly and the average angle values increases to values close to 90 degrees (i.e. completely orthogonal prediction vectors).

Even under extreme operating conditions with voltage integrity attacks carried out using our strongest attacker circuit from Section \ref{sec:attack_char}, we can still achieve 18\% higher inference performance with no effect on the prediction accuracy by overclocking the DL accelerator. 
If we increase the clock frequency beyond 680 MHz and 650 MHz in the case of the clock-gated garbled XORs and clock-gated hybrid logic attackers respectively, the accelerator is not able to finish the classification of all 50,000 inputs as some paths in the control logic start to fail timing.
To decide the degree of safe overclocking at runtime, a few timing violation detection circuits (shown in Fig. \ref{fig:victim}a) can be sprinkled at different locations of the FPGA role to detect voltage drops and dynamically adjust the operating frequency of the DL accelerator.
This approach would result in unaltered predictions in addition to a 1.18-1.3$\times$ performance boost at the cost of less than 1\% more resources. 

\subsubsection*{\textbf{Model Resilience}}
The plots in Fig. \ref{fig:hpipe} also highlight the inherent resilience of the DL model against induced timing faults.
For instance, at 690 MHz and 680-710 MHz frequencies in Fig. \ref{fig:hpipe}a and \ref{fig:hpipe}b respectively, although the prediction vectors produced by the DL accelerator are different as indicated by the non-zero average angle values, the overall prediction accuracy remains unaffected.
Therefore, the accelerator's operating frequency can be viewed as an additional knob that offers a performance/accuracy tradeoff.
All our experiments in this study are conducted on a dense model using 16-bit fixed point precision.
However, we believe that studying the models' resilience to timing faults at different numerical precisions and degrees of sparsity is an interesting future work.

\vspace{-0.1cm}
\section{Conclusion}
FPGAs are being widely deployed in datacenters and are envisioned to become a virtualized multi-tenant compute platform in the near future.
However, this comes with potential security threats arising from the reconfigurable nature of FPGAs.
In this study, we focus on integrity attacks in a multi-tenant FPGA scenario.
An adversary can cause voltage drops by implementing a malicious circuit in her role, inducing timing violations in the victim role, even when physically isolated. 
We demonstrate the feasibility of such integrity attacks with three different adversarial circuits, two of which utilize a vendor-supplied clock-gating IP core. 
An attacker that combines hard and soft logic with clock gating is as effective as a traditional RO attacker, but will not be detected by bitstream verifiers. 
We show that while floorplanning the attacker and victim circuits to physically separated roles has some positive effect, the integrity attacks still succeed. 
Finally, we carry out an integrity attack on a DL accelerator running ImageNet classification in the victim role to study the resilience of DL models against induced timing faults.
We show that, due to this inherent resilience, safe overclocking of the DL accelerator can result in 1.3$\times$ and 1.18$\times$ higher performance under normal operating conditions and extreme voltage attacks, respectively.
These results indicate that current FPGAs are vulnerable to voltage integrity attacks, creating a barrier to multi-tenant applications and raising the possibility of inadvertent failures within a single design due to aggressive clock gating. On the positive side, DL inference shows significant resilience to timing faults, giving additional resistance to voltage integrity attacks and allowing performance increase via safe overclocking.

\vspace{-0.1cm}
\section*{Acknowledgements}
\vspace{-0.1cm}
The authors would like to thank Ibrahim Ahmed for the insightful discussions, and the NSERC/Intel industrial research chair in programmable silicon and the Vector institute for funding support.

\bibliographystyle{./bibliography/IEEEtran}
\bibliography{ref.bib}

\end{document}